# DesignCon 2006

# Time Domain Verification of Differential Transmission Line Modeling Methods

Jonathan D. Coker, Mayo Clinic

Dr. Erik S. Daniel, Mayo Clinic

Dr. Barry K. Gilbert, Mayo Clinic
gilbert.barry@mayo.edu   507-284-4056

## Abstract
The advantages and limitations of time-domain pseudo-random binary sequence (PRBS) excitation methods for system identification of individual modes within a multi-conductor transmission system are discussed. We develop the modifications necessary to standard frequency-domain transmission-line models to match time-domain experimental data from several types of transmission systems. We show a variety of experimental results showing very good to excellent agreement with our model's predictions, up to approximately 10 GHz.

## Author Biographies

Jon Coker is a Principal Project Engineer of the Special Purpose Processor Development Group at the Mayo Clinic. Mr. Coker graduated from Wheaton College with a Bachelor of Arts degree in 1982 and from the University of Minnesota with a Bachelor of Science degree in electrical engineering in 1984. He is currently pursuing a Ph.D. degree at the University of Minnesota.

Erik Daniel received the B. A. degree in physics and mathematics from Rice University, Houston, TX, in 1992. He received the Ph.D. degree in solid state physics from the California Institute of Technology, Pasadena, CA, in 1997, with thesis research focusing on simulation, fabrication, and characterization of quantum effect semiconductor devices. He currently is a Staff Scientist in the Department of Physiology and Biomedical engineering, Mayo Clinic, Rochester, MN, and the Deputy Director of the Special-Purpose Processor Development Group.

Barry Gilbert received the B.S. degree in electrical engineering from Purdue University, West Lafayette, IN, in 1965 and the Ph.D. degree in physiology and biophysics (with minors in electrical engineering and applied mathematics) from the University of Minnesota in 1972. He is currently a Staff Scientist in the Department of Physiology and Biomedical engineering, Mayo Clinic, Rochester, MN, and the Director of the Special Purpose Processor Development Group.

# Introduction

High-speed digital back-plane communication channels have long since given up their "digital design" status and have come to resemble (architecturally) previously-existing, complex analog communication systems.  Equalization, error correction, modulation codes, or advanced detection schemes are now commonly proposed and implemented for serializer-deserializer (SERDES) channels (see [10], for example). To achieve the maximum benefit of these methods,  we require increasingly accurate system identification of the transmission system.  At the same time,  we find that traditionally excellent transmission-line models begin to diverge from experiment as bandwidths begin to extend into the GHz and tens of GHz range [13]. Thus reliable methods for system identification and model verification for candidate transmission systems supporting such channel implementations are of increasing and fundamental importance.

Several methods exist for accurate system identification of linear systems.  Analysis using network analyzers (including vector network analyzers) is a highly-developed technology using frequency-domain measurements.  Advanced time-domain methods have more recently come on the scene employing TDR and TDT measurements of a step excitation [7-9].

In this paper, we present a technique to predict and experimentally verify transmission line models purely in the time domain using the method of pseudorandom sequences.  Our technique is an extension of a method developed for system identification commonly used in magnetic recording data channels and testers [3,6].    We shall focus on the description of transmission lines which are suitable for a high-speed serial-communication link.  We shall compare expected and experimental results from printed circuit board (PCB) transmission lines and to results from cables.   While we shall not argue that the proposed method is experimentally superior to established time-domain techniques for linear systems,  we shall discuss the unique nonlinear-system-identification capabilities of a PRBS waveform.  In addition, we shall show a fast algorithm which is simple enough to consider implementing in modern SERDES hardware, thus allowing a wide range of inexpensive and highly capable built-in self-test and board diagnostic capabilities.

We shall describe our specific model of a transmission line.   This section is intended to outline the assumptions and limitations of the model.   We also present our alternative extensions to the standard telegrapher's model, which we found necessary to adequately describe experiment in some cases.

# Transmission Line Modeling
In this section, we briefly review a classical transmission-line model, and discuss the parameters which we use in the model.
## Telegrapher's Model in the Time Domain

The basic telegrapher's model for transmission lines is fully worked out in several texts [1,2]. The standard solution gives the voltage transfer function for a wave traveling in the positive $z$ direction as

$$H(\omega) = e^{-\gamma(\omega)\Delta z} \tag{1.1}$$

where $\omega$ is the radial frequency, $\Delta z$ is the distance down the line, and the complex propagation constant $\gamma(\omega)$ is:

$$\gamma(\omega) = \sqrt{(R+j\omega L)(G+j\omega C)} = \sqrt{(R+sL)(G+sC)} \tag{1.2}$$

If the Fourier transform of an input waveform $x(t)$ is $X(\omega)$, then the Fourier transform of the waveform at the output of the transmission line, $y(t)$, is

$$Y(\omega) = H(\omega)X(\omega) \tag{1.3}$$

and the time-domain version of the output waveform is

$$y(t) = \mathrm{F}^{-1}\{H(\omega)X(\omega)\} \tag{1.4}$$

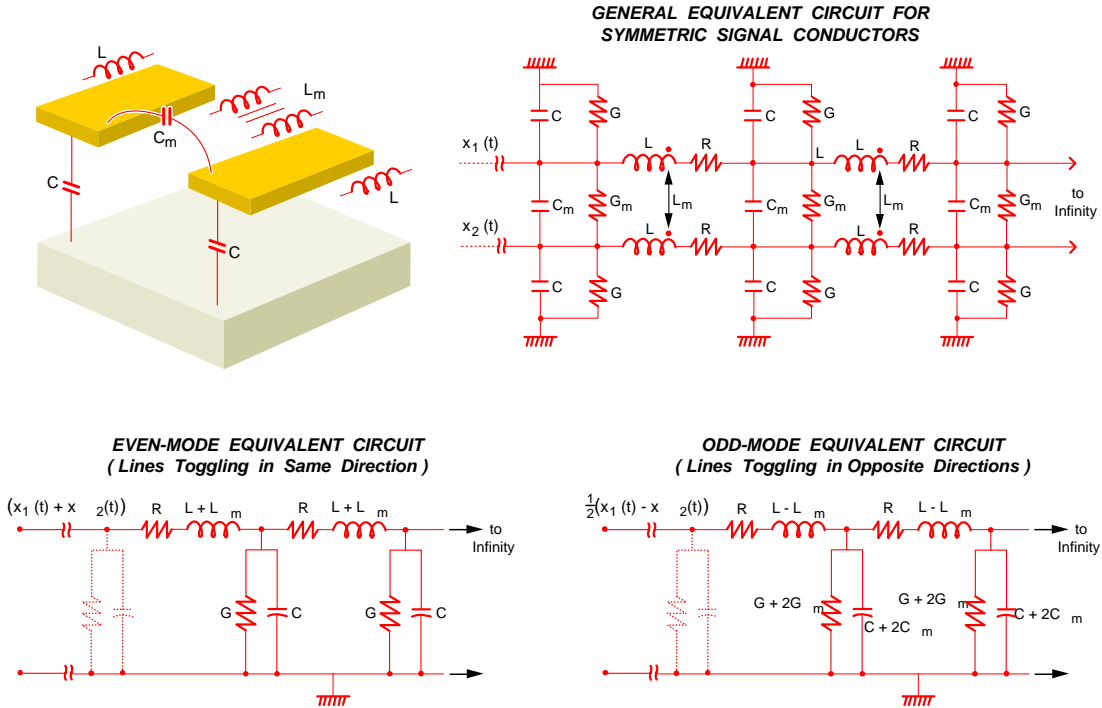

**Figure 1: Simplified Odd Mode and Even Mode Equivalent Circuits for a Symmetric Two-Signal-Conductor Generalized Transmission Line (18500).**

In Figure 1, we show the standard, generalized equivalent circuit of a symmetric two-signal-conductor transmission line, and reduced-complexity, single-ended equivalent circuits for each transmission mode. Note that the simplified equivalent circuits allow direct application of the RLGC model, using the correctly-transformed versions of the RLGC parameters appropriate for the transmission mode.

## Variation of RLGC Parameters with Frequency

Having set the stage for the time-domain application of a generic RLGC model, we now focus on the specific forms of each component used our RLGC model. The following section specifies the formulas used in our basic RLGC model (which are fairly standard) and identifies modifications we thought necessary to adequately explain observed laboratory behavior (which are not always standard).

## Series Impedance Variation with Frequency

In the case of simple, homogeneous signal conductors, we use the classical result for the series impedance based on surface impedance concepts, which we repeat here:

$$(R + sL) = R_{AC} \sqrt{s} + L_{\infty} s \tag{1.5}$$

where $L_{\infty}$ may be interpreted as the inductance of the system when all currents flow uniformly on the surface of the signal conductors (that is, at moderately high frequency) and $R_{AC}$ is a constant which we approximate as

$$R_{AC} = \frac{\eta}{S} \sqrt{\frac{\mu}{2\sigma}} \tag{1.6}$$

where $\mu$ and $\sigma$ are the permeability and conductivity of the signal conductor, and $S$ is the length of the effective perimeter of the signal conductor through which the surface currents flow. The geometry-dependent constant $\eta$ represents a factor determining the increase in resistive losses due to currents in the return paths. In a thin stripline configuration, one might expect the value of $\eta$ to be in the neighborhood of $\eta = 2$ because the widths of the expected return paths are about the same as the circumference of the signal conductor. In a coaxial cable, one might expect $\eta$ to be less than 2, because the return paths in the outer shield are significantly wider than the circumference of the signal conductor. In practice, any of the parameters of equation (1.6), including $R_{AC}$ itself, may be varied in equation (1.5) to match laboratory data.

However, there exists a common transmission system which does not fit equation (1.5) well. The Gore EyeOpener ™ cable, for example, uses signal conductors constructed from a heterogeneous combination of metal layers to achieve self-equalizing properties. We now derive an approximation to the surface impedance for a thick bulk material, covered by a relatively thin layer of another conductor, to address this case. General field solutions to this type of problem were generated by Wait [5], which we here apply to transmission lines.

We presume a planar, infinitely-thick conductor of bulk conductivity $\sigma_2$ underneath a thin layer of thickness $\tau_1$ and conductivity $\sigma_1$. As in the classical case, the electric field will decay throughout the finite thickness $\tau_1$ to the value

$$E_z(y)|_{y=\tau_1} = E_0 e^{-\frac{1+j}{\delta_1}\tau_1} \qquad (1.7)$$

where $\delta_1$ is the skin depth in the outer conductor. At the interface between the two different conductors, the electric field will have a new behavior given by the boundary condition requirements of Maxwell's equations. The appropriate constraint is that of continuous tangential electric field across the interface. Therefore, at the interface, the electric field begins a new exponential behavior in the bulk material with new depth constant $\delta_2$:

$$E_z(y) = E_0 e^{-\frac{1+j}{\delta_1}\tau_1} e^{-\frac{1+j}{\delta_2}(y-\tau_1)} \qquad (1.8)$$

Working out the integral for the total current under a width $S$, the resulting surface impedance is:

$$\begin{aligned}Z_z &= \{\frac{1+j}{S\sigma_1\delta_1}\}\{e^{-\frac{1+j}{\delta_1}\tau_1}(\frac{\sigma_2}{\sigma_1}-1)+1\}^{-1} \\ &= \{R_{AC_1}\sqrt{s}\}\{e^{-\eta\sqrt{\frac{\mu_1 s}{2\sigma_1}}}(\frac{\sigma_2}{\sigma_1}-1)+1\}^{-1}\end{aligned} \qquad (1.9)$$

Equation (1.9) adds a new factor to the classical surface impedance. The factor is a function of the outer-conductor thickness $\tau_1$ and the ratio of the conductivities of the two materials. In general, we see that the resistive and reactive portions of the surface impedance are no longer equal when the conductor is composite. In our physical approximation, we take the conductor width $S$ to be the effective length of the perimeter of the signal conductor. The overall series impedance is then:

$$R + sL = \{R_{AC_1}\sqrt{s}\}\{e^{-\eta\tau_1\sqrt{\frac{\mu_1 s}{2\sigma_1}}}(\frac{\sigma_2}{\sigma_1}-1)+1\}^{-1} + sL_\infty \qquad (1.10)$$

Equation (1.9) is valid when the permeability of the all conductors is that of free space. When the bulk conductor is magnetic, the conductivity of the bulk conductor $\sigma_2$ can be replaced by an effective conductivity

$$\sigma'_2 = \frac{\sigma_2}{\mu_R} \qquad (1.11)$$

where $\mu_R$ is the relative permeability of the bulk conductor.

## Shunt Conductance Variation with Frequency

Traditionally, the dielectric losses are characterized by a shunt conductance *G* per unit length:

$$G = \omega C \tan \delta = \frac{1}{j} sC \tan \delta \qquad (1.12)$$

The "loss tangent", $\tan \delta$, is commonly specified by dielectric manufacturers to be in the 0.01 to 0.001 range. Typically, users are left to presume that the loss tangent is independent of frequency. In such a case, the total shunt admittance can be written as

$$G + Cs = \omega G \tan \delta + Cs = (1 - j \tan \delta)Cs \qquad (1.13)$$

We shall see that this form can give good agreement with laboratory data at frequencies in the few-GHz range (and when dielectric losses are relatively small compared to the series resistance). At higher frequencies (perhaps in the 10 GHz range) the model's main deficiency becomes apparent: the form of equation (1.13) cannot be physically reasonable. It is well known that this form of dielectric loss is not a physically consistent possibility [14]. In the present case, it is relatively straightforward to see why this is so. If we take the series impedance to be lossless, that is, $R_{AC} = 0$, then using equations (1.13), (1.5), (1.2), and (1.1), the transform of the impulse response of the transmission line can be written as:

$$\mathrm{F}(h(t)) = e^{-\gamma \Delta z} = e^{-j\omega \sqrt{LC(1 - j \tan \delta)}} \qquad (1.14)$$

Equation (1.14) has a closed-form inverse transform, which is:

$$h(t) = \frac{\alpha}{\pi[(t - \tau)^2 + \alpha^2]} \qquad (1.15)$$

where the parameters are given as

$$\begin{aligned} \tau &= \Delta z \sqrt{LC} \, \mathrm{Re}\{\sqrt{1 - j \tan \delta}\} \\ \alpha &= \Delta z \sqrt{LC} \, \mathrm{Im}\{\sqrt{1 - j \tan \delta}\} \end{aligned} \qquad (1.16)$$

Therefore, the impulse response of a transmission line with dielectric possessing constant loss tangent is a delayed Lorentzian pulse. The Lorentzian form gives experimentally plausible insights, such as: the amplitude of the pulse is inversely proportional to the length of the transmission line, with proportionality constant simply related to the loss tangent. However, we can observe that the impulse response extends back infinitely in time even though its impulse excitation occurred at the time origin. The model predicts non-causal behavior and is therefore not plausible as a physical model. We have found that this feature of the constant-loss-tangent model is often the root cause of the failure to match our experimental data at high frequency.

Other forms for the variation of the loss tangent with frequency must be applied in these cases. In Figure 2 we highlight the difference in expected pulse shape between the constant-loss-tangent model and that of a loss tangent varying linearly with frequency. The response in the linear-loss-tangent case is derived numerically using equation (1.4) because the problem does not have a closed-form solution. We will show later that the linear-variation version can exhibit good fit to experimental data at frequencies in excess of 10 GHz (which is our only justification for using it).

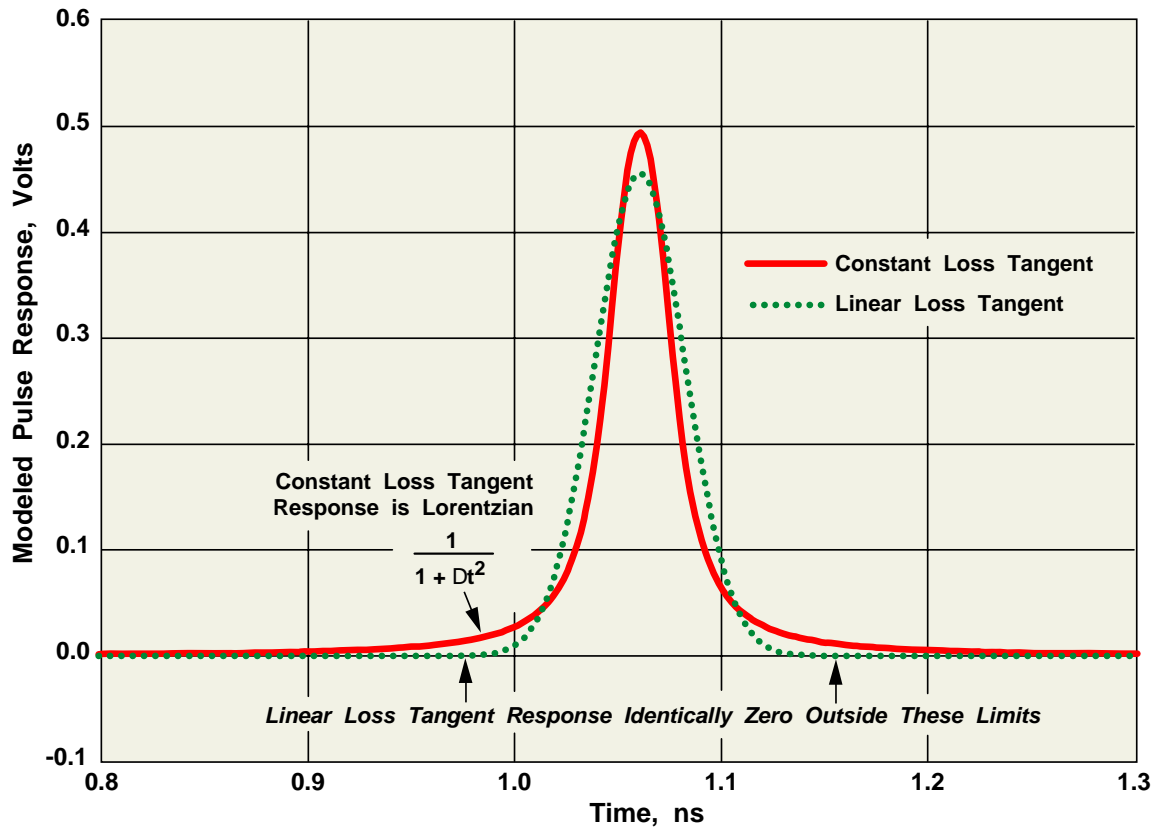

**Figure 2: Comparison of Modeled Time Responses of a Transmission Line with No Series Loss but Finite Dielectric Loss Due to Two Loss Models, Showing Causality Failure of Constant Loss Tangent Assumptions. (20573)**

## Time-Domain Laboratory Techniques

Transmission lines (and other linear circuits) are often characterized by s-parameter analysis using a vector network analyzer (VNA). The methodology for calibration, de-embedding, and interpretation of VNA results is a highly-developed specialty, which (when properly applied) will fully characterize the transmission line over a wide bandwidth.

For this work, we have elected to use a time-domain method, for the following reasons. First, most VNAs have two ports and do not simply support differential-mode excitation.

Second, in many applications, the primary information needed is the classical transfer-function of the transmission line system (i.e., the information present in $s_{21}$ in the absence of reflections). The complexities in testing, calibration, de-embedding, and interpretation for the other s-parameters may not be strictly necessary in some applications. Third, a full-fledged experimental characterization of systems utilizing transmission lines is often not limited to pure system-identification techniques, but also may include direct measurements of higher-level system performance quantities (such as error rate or eye diagrams). In some applications it is beneficial to integrate as many measurement schemes as feasible into an experimental setup that is as simple (and inexpensive) as possible, so long as the resulting measurements are "good enough" for the application.

Finally, there may exist nonlinearities in the transmission system which will introduce interpretation errors in the s-parameter analysis without warning. These nonlinearities are not likely to be in the transmission lines and interconnect (which are linear to an excellent approximation) but may exist, for example, in the driver circuitry of a buffer amplifier. Most (if not all) nonlinearities cannot be characterized by the magnitude and phase of single-sinusoid reception; however, there exist waveforms which do broaden the scope of complete nonlinear characterization. We shall see that a PRBS is one such waveform.

In the following section, we describe a method which addresses all four of these points. Naturally, the described method will have its own disadvantages, which are generally related to experimental approximations which do not exist (or are unnecessary) in a VNA-type of analysis.

## Pseudorandom Sequence Excitation Method

We describe a time-domain system identification method in this section. We follow the scheme developed in [3,6]. A performance analyzer ('bit error rate tester') is used to generate a 127-bit pseudo-random binary sequence (PRBS). The analyzer has true and complement outputs, which are used to drive the differential transmission line. A digital oscilloscope is used to sample the differential signal at the end of the transmission line. The scope is triggered by the performance-analyzer synch-out pulse, which fires once every PRBS period. This pulse provides a consistent time datum which is independent of the transmission-line length. A typical experimental setup (shown for a cable) is as shown in Figure 3.

The upper panel in Figure 4 shows an example of the raw waveform as sampled at the end of a transmission line. Because the raw waveform is too complicated to interpret easily by eye, the waveform can be processed to generate a discrete-time pulse response.

15 Meter Cable Assembly Eye Diagram Test

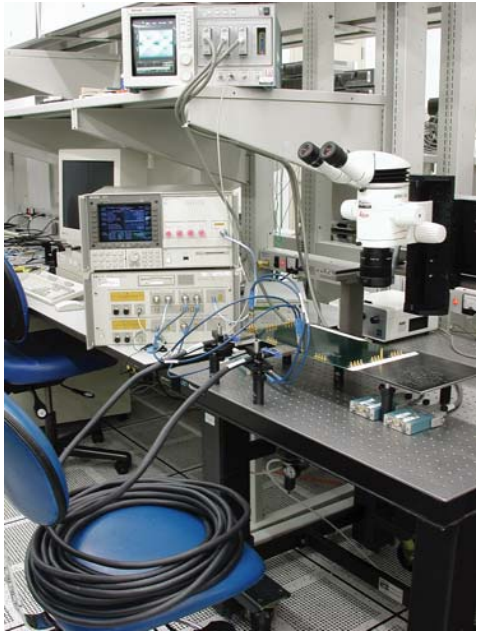

Board With Cable Assembly Under Test

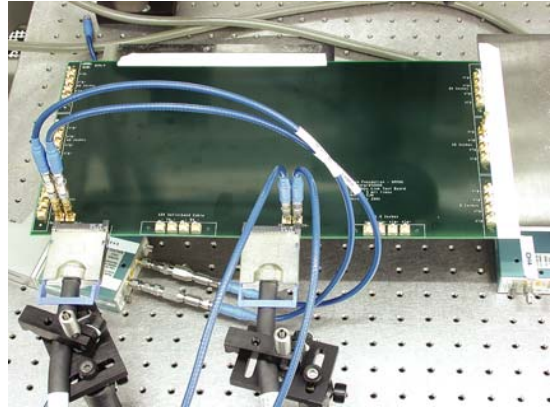

**Figure 3: Test Setup for Measuring Error Rate, Eye Diagrams, and System Transfer Functions in the Time Domain. (18045)**

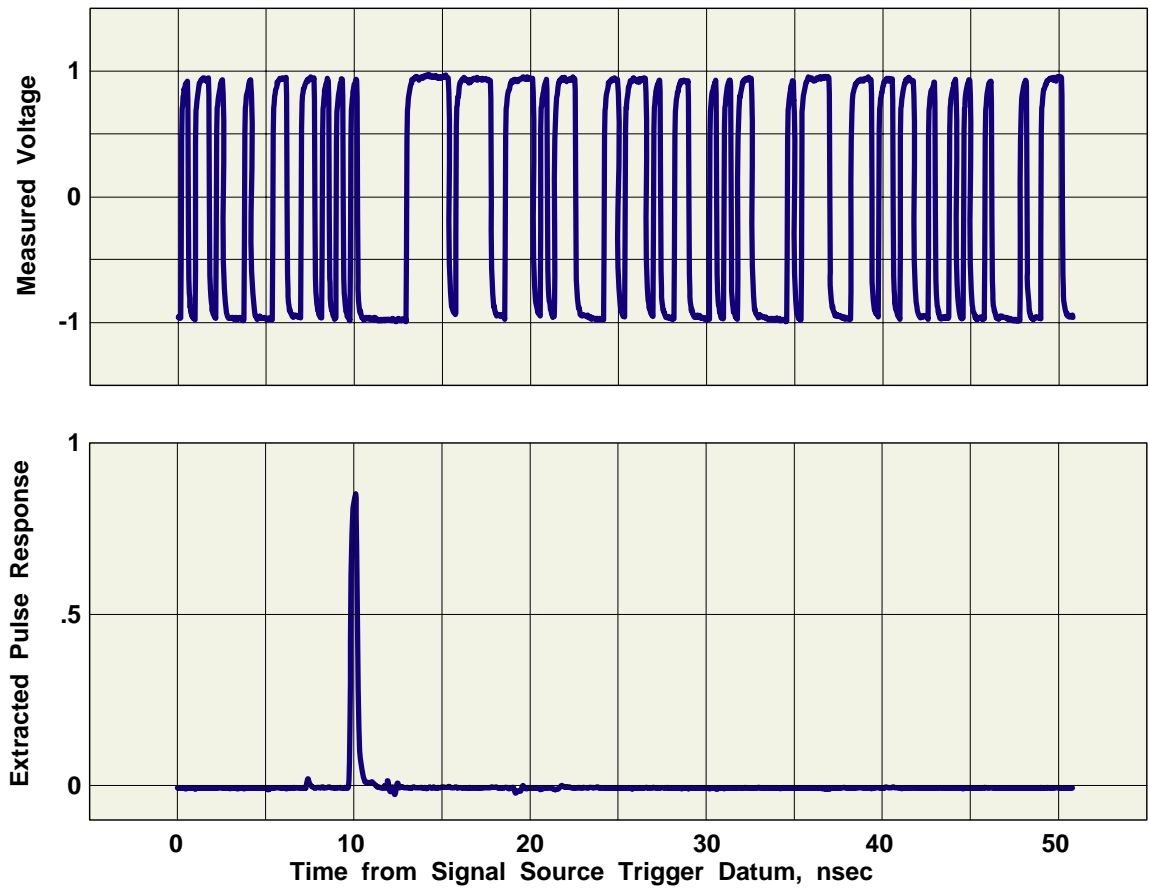

**Figure 4: Example of a Measured PRBS Waveform and Its Extracted Pulse Response. (18322)**

The response (which we shall call *extracted pulse response* in keeping with the literature [3,6]), is numerically generated using a two-step process. First, the raw waveform is re-sampled to take care of experimental frequency errors between the PRBS generator and the oscilloscope, and to provide an appropriate sampling rate for subsequent analysis. Second, the re-sampled waveform is mathematically deconvolved with a theoretically perfect version of the same PRBS. We shall return to describe a very efficient method for this deconvolution.

The extracted pulse response is analogous to the impulse response in continuous time systems. In this case, the extracted response is the expected response of the transmission line (and measurement system and signal drivers) if the PRBS driver had output a simple base-to-peak digital pulse instead of the PRBS.

The lower panel in Figure 4 shows the extracted response for the given raw waveform. While both waveforms contain exactly the same information, it is clear that the extracted pulse is a short primary response, with some small echoes of interesting shape and location on the time axis. In Figure 5, the extracted pulse response for two different, but short, PCB transmission lines is shown. All connectors, test cables, and test conditions are nominally identical, except for the length of the transmission line.

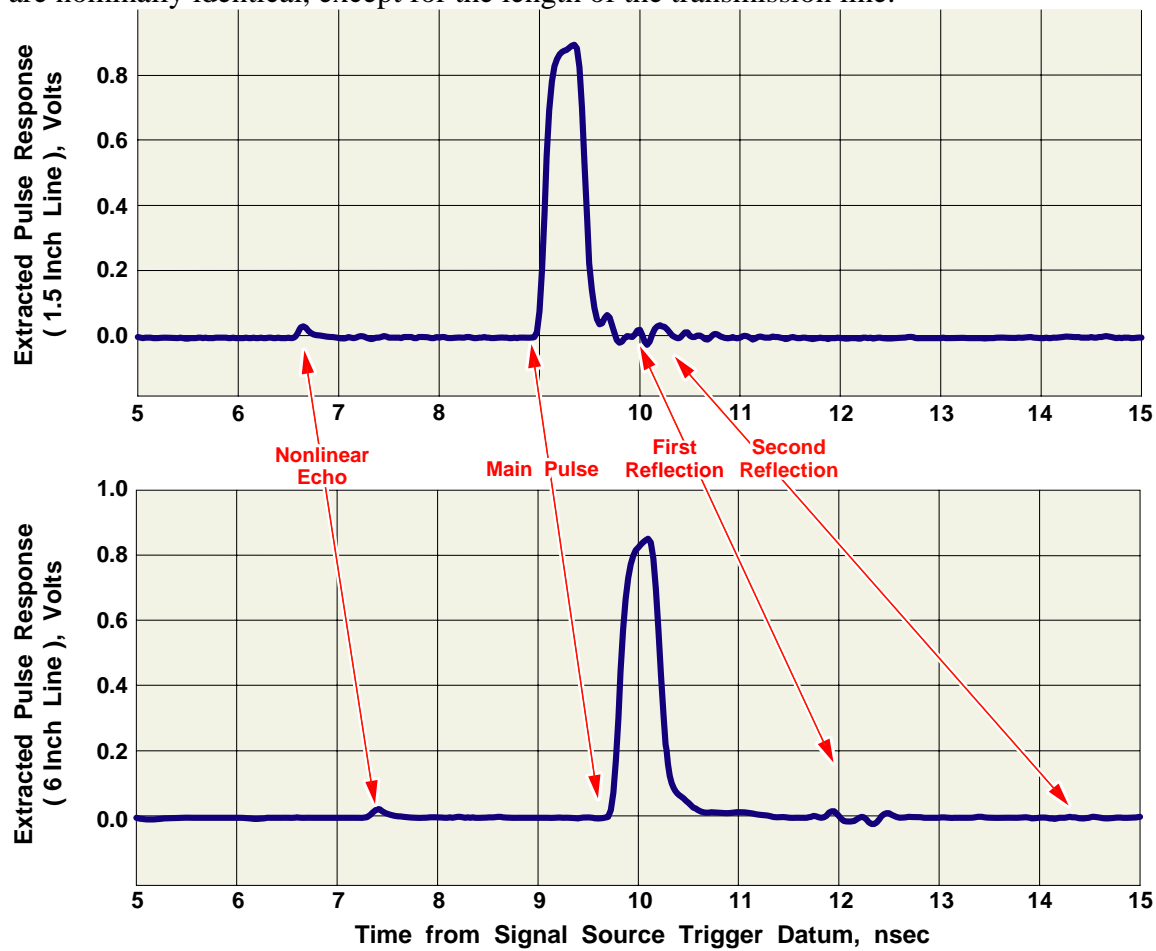

Figure 5: Extracted Pulse Responses from Two Lengths of Otherwise Identical Lines (1.5" and 6") (18323)

In the upper panel of Figure 5, we zoom in on the interesting sections of the extracted pulse response, for a 1.5 inch PCB line. As might be expected from the significant impedance mismatches, the signal appears to reflect back and forth between the impedance discontinuities before settling down. At 2.5 Gbits/second excitation, the pulse response is nominally 400 picoseconds wide, and the first forward-moving reflection should begin to arrive about a half nanosecond after the onset of the main pulse; the second forward-moving reflection arrives a half nanosecond after that, and so forth. Thus we should not expect to see separation between the main pulse and its subsequent, exponentially decaying reflections.

In the lower panel of Figure 5, we see the effects of quadrupling the transmission-line length. The now well-resolved reflections occur at approximately 2 nanosecond intervals (as expected). The main pulse is resolved, and is proportional in magnitude to what the pulse would have been down 6 inches of this transmission line *without the reflections due to impedance mismatches*, but including any pulse modification due directly to impedance mismatches. (The interface of two perfect transmission lines of real, but different, impedance at all frequencies will modify the main pulse in amplitude, only).

The main pulse and reflections may also be resolved by increasing the excitation frequency of the PRBS on a given line (instead of changing line lengths, as above). This technique may be limited by the rise time of the drivers, the time span of the impulse response of the transmission lines under test, and the analog bandwidth of the sampling oscilloscope.

These observations suggest a type of de-embedding technique which is very simple, though approximate. We assume that the combination of the transmission line length, the resolution of the channel response, and the frequency of the PRBS is such that the main pulse is resolvable from reflections or other imperfections. We then claim, to the extent these assumptions are true, that the resolvable main pulse may be written as

$$\begin{aligned}x_1(t) &\cong F^{-1}\{X_0(\omega)H_{driver}(\omega)H_{launches}(\omega)H_{scope}(\omega)...H_{tline}(\omega,\Delta z_1)\} \\ &\cong F^{-1}\{X_0(\omega)H_T(\omega)H_{tline}(\omega,\Delta z_1)\}\end{aligned} \qquad (1.17)$$

where $X_0(\omega)$ is the Fourier transform of an ideal PRBS sequence, $\Delta z_1$ is the length of transmission line, and the total transfer function $H_T(\omega)$ is the composite linear response of all non-ideal elements in a practical test. Note that the interpretation of $x_1(t)$ is not exactly that of the time response related to the s-parameter $s_{21}$; it is only equivalent to s-parameter analysis when there are no reflections. In principle, and with sufficient cleverness, we could also resolve each reflection in the measured waveforms to independently and fully analyze the entire characteristic described by $s_{21}$. However, we shall not develop those techniques in this report; instead, we shall focus only on resolving the response that would have occurred without reflections.

If we now consider a transmission line of length $\Delta z_2 > \Delta z_1$, then its response can be written as

$$\begin{aligned} x_2(t) &= \mathrm{F}^{-1}\{X_0(\omega)H_T(\omega)H_{tline}(\omega, \Delta z_2)\} \\ &= \mathrm{F}^{-1}\{X_0(\omega)X_1(\omega)H_{tline}(\omega, \Delta z_2 - \Delta z_1)\} \end{aligned} \quad (1.18)$$

where the second line follows from the separability of the transmission line transfer function (equation (1.1)) into component factors, that is,

$$\begin{aligned} H_{tline}(\omega, \Delta z_2) &= e^{-\gamma(\Delta z_2)} = e^{-\gamma(\omega)(\Delta z_1 + \Delta z_2 - \Delta z_1)} = e^{-\gamma(\omega)(\Delta z_2 - \Delta z_1)}e^{-\gamma(\omega)(\Delta z_2 - \Delta z_1)} \\ &= H_{tline}(\omega, \Delta z_1)H_{tline}(\omega, \Delta z_2 - \Delta z_1) \end{aligned} \quad (1.19)$$

Despite its mathematical look, the suggested modeling technique is exceedingly simple:

1. Measure $x_1(t)$ of a relatively short line and find its extracted pulse response.
2. Calculate the expected response of a longer transmission line of length $\Delta z_2$ by applying equation (1.4) for a transmission line of length $\Delta z_2 - \Delta z_1$.
3. Compare the output of step 2 to an actual measurement of the extracted pulse response of a transmission line of length $\Delta z_2$. Focus on the (by assumption, *resolvable*) main pulse only; ignoring *all other interesting blips.*

We have argued that this technique will give an expected waveform, and an actual waveform, to measure the goodness-of-fit of the RLGC model; furthermore this comparison de-embeds all systematic linear effects in the test, to the extent that the main pulse is resolvable in time from all reflections or other phenomenon.

The "resolvability" criterion gives this technique its simplicity, but is its major source of interpretation uncertainty when compared to more accurate experimental methods (such as, a VNA). A typical "eyeball" comparison is good to perhaps 30-40 dB (a few percent in voltage). In many applications (such as, the identification of eye diagrams), the "eyeball" criterion is, nearly by definition, good enough. However, the engineer must judge what kind of accuracy is required for a particular application.

We have argued that the proposed technique effectively de-embeds effects which can be described by a linear transfer function. We now briefly digress to discuss the effects of stochastic jitter in the trigger waveform and oscilloscope timings. These effects are not present in frequency-domain measurement techniques, and therefore deserve further discussion.

In general the analysis of the all jitter effects on the waveform captured by the oscilloscope is very difficult, but the following simplifications allow some insight into the problem. If the jitter is small enough that the waveform is well approximated by its first-order Taylor's series, that is:

$$x(t + \Delta\tau) \cong x(t) + \Delta\tau x'(t) \qquad (1.20)$$

and all jitter phenomenon are lumped into a single effective jitter at the scope sampler (relative to the signal $x(t)$), with jitter variance $\sigma^2$, then the signal can be thought of in two components. There exists a stochastic portion (whose spectrum is continuous even if $x(t)$ is periodic) whose power is linearly related to $\sigma^2$ and the average square of the derivative of $x(t)$. This power comes at the expense of the high frequencies in the deterministic portion of the identified waveform (that is, its averaged sequence value as sampled on the scope). It can be shown that the jitter produces a stochastic filter whose average transfer characteristic is

$$\tilde{H}_{jitt}(\omega) = e^{-\sigma^2 \omega^2} \qquad (1.21)$$

if the jitter is Gaussian. Therefore, if $f \ll 2\pi/\sigma$, the jitter effects should be negligible in the sampled waveform $x(t)$. If the standard deviation of the jitter is in the few pS range, then frequencies below about 10 GHz are affected by less than 1 percent.

In a sense, $\tilde{H}_{jitt}(\omega)$ can be thought of as one of components of the total deterministic transfer function $H_T(\omega)$, but only if the various realizations of $\tilde{H}_{jitt}(\omega)$ are averaged sufficiently to converge to the average transfer function. This convergence might happen, for instance, when the waveform reported from the scope is the average of a large number of sample sequences. In such a case, this portion of the transfer function cancels out in our method (as do the deterministic transfer functions). However, this is dangerous ground, and the experimenter should take care to ensure that the stochastic conditions of the tests are identical when exploiting this effect. For example, if trigger jitter is significant, the realization of $\tilde{H}_{jitt}(\omega)$ for a waveform derived by "averaging" only one sample waveform is in general quite different from the realization of $\tilde{H}_{jitt}(\omega)$ when many averages are taken. (The frequency response will look better, in general, with only a few averages than it does with long averages).

### Numerical Methods

In the preceding section, we developed the concept of the extracted pulse response, which is the circular deconvolution of the measured PRBS sequence with an ideal version of the same sequence. A good, standard method for deconvolution is to take the Fourier transform of both waveforms, perform a complex division, and then take the inverse Fourier transform.

In the present case, the number of points in the transform is $n = 2^N - 1$, which is very often a prime number, and is never highly composite. In such cases, the best fast algorithms for discrete Fourier transforms (DFTs) are relatively useless. In most such cases, the deconvolution will require about $n^2$ complex multiplication operations and about $n$ complex divisions.

We now show a special deconvolution method which takes optimal advantage of the fact that measured waveform is a filtered version of a PRBS. This method requires about $n^2$ real additions, zero multiplications, zero divisions, and zero complex operations of any sort.

We develop the algorithm in the following way. If the sampled digital-pulse response of the transmission line system is denoted by $h[n] = h(nT)$, with $n$ an integer and $T$ the bit period of the sequence, then the expected sampled measured output is giving by the circular convolution of the PRBS and the pulse response:

$$y[n] = \sum_k h[k-n]p[k] \qquad (1.22)$$

where $p[n] \in \{-1, 1\}$ represents the PRBS and $k$ ranges over the length $L$ of the PRBS, which is always of the form $L = 2^N - 1$. We shall also call $h[n]$ the *extracted pulse response*, after we have back-calculated it from measured data. Because this is a circular convolution, the sequences $y$, $p$, and $h$ are assumed to be periodic in $L$; therefore the indices can always be taken modulo $L$.

We now assume that the mathematical convolution was actually completed in the physical world by playing the ideal PRBS through a linear system, and we are able to measure the analog waveform $y(t)$ directly. We then resample this waveform to obtain the measured version of $y[n]$. By inspection, we can generate a new function from $y[n]$ and note how else it can be expressed:

$$\begin{aligned} Z[n] &= \frac{1}{L+1} \sum_l y[n]p[n-l] \\ &= \frac{1}{L+1} \sum_l \sum_k h[k-n]p[k]p[n-l] \\ &= h[n] + \mu_x \end{aligned} \qquad (1.23)$$

where $\mu_x$ is the mean of the pulse response $x[n]$. The last line of equation (1.23) is general only when the excitation is a PRBS, and is derived using the properties of a PRBS [4]. Therefore we can recover, within a small DC shift, the extracted pulse response $h[n]$ from the measured sequence $y[n]$ by a simple transformation using $L(L-1) \cong L^2$ additions and $L$ trivial multiplications by $\pm 1$. Depending on the application, the division by L+1 may or may not be necessary; if it is necessary, implementers should note that the divisor is always exactly a power of two. With a little more complexity in the algorithm it is possible to reconstruct exactly the DC conditions to get a new function $Z'[n] \equiv h[n]$, but in our case we have simply ignored the small DC offset as it makes little practical difference in our analysis.

We have developed this algorithm in a general way when we are interested in analyzing or plotting waveforms with more than one point per PRBS bit. In such a case, we can decimate a waveform representing $M$ points per bit into $M$ waveforms representing $y_m[n]$, that is, the sampled version of $y((n+m/M)T)$. We can determine the extracted pulse response for each sub-channel, $h_m[n]$, using equation (1.23) and $y_m[n]$ as input. The resultant response $h_m[n]$ is the sampled version of the analog waveform $h((n+m/M)T)$. Therefore, we can reconstruct the $M$-point-per-bit version of the response by re-interleaving the results of each $h_m[n]$. Note that the calculations of $h_m[n]$ are *only* dependent on values of $y$ in its own interleave; whereas in general this property is not true of DFT-type deconvolutions. This property is another advantage of this type of specialized deconvolution.

Therefore, we have shown a very efficient method for deconvolution when PRBS excitation is used. Clearly, for offline analysis using relatively small $L$, there is little practical difference between a few microseconds and a several milliseconds of computation time. Even so, these differences can become quite significant for large enough $L$ even for offline analysis because $L$ grows exponentially with each PRBS order $N$. Perhaps more significantly, the algorithm in equation (1.23) is simple enough to realistically be implemented at-speed in hardware. For example, the algorithm implementation is simply a single accumulator if we are willing to calculate one $Z(n)$ at a time in an operation.

## Nonlinear Analysis with PRBS Excitation

Up to this point, we have not described anything of particular value in using a PRBS as the excitation waveform over other types of waveforms (other than its alternative utility as a convenient bit-error-rate test pattern). Indeed, a TDT-type step waveform gives the same information as does an extracted pulse response for a linear system. We now digress to explain an interesting and useful feature of this type of PRBS-deconvolution analysis, which allows native nonlinear Volterra analysis of the waveforms.

In each of the extracted pulse waveforms discussed thus far (shown in Figure 4 and Figure 6), a strange blip appears to arrive at the oscilloscope 2.5 nanoseconds *before* the main pulse. This blip always shows up in exactly the same place (relative to the main pulse), independent of the transmission-line length or type, or even whether there is a transmission line under test at all. If the bit excitation frequency is changed, then the time spacing between the blip and the main pulse also changes. However, if the time spacing is normalized to bit time units, now *this* number now remains constant, independent of excitation frequency. Therefore, we have discovered an apparently physical delay effect which seems to know something about the bit times. No signal phenomenon of any kind shows up at these locations when a simple TDR pulse is used as excitation.

The reason for this type of behavior was resolved in the 1980s [3]. The phenomenon came to be understood in the following way. If a system is linear, then the deconvolution operation, described above, always gives the same results for the extracted pulse response no matter what the underlying data pattern is. If a system is somewhat nonlinear, then in the general case, the deconvolution operation will contain garbage which will appear as a "noise" everywhere on the extracted pulse response. However, if the data pattern is a PRBS, then for many types of nonlinearities, the disturbances due to such nonlinearities destructively interfere almost everywhere after the deconvolution operation. The nonlinearities tend to constructively interfere at specific locations on the extracted pulse response, and with specific shapes and magnitude, all of which vary depending on the nature and severity of the nonlinearity. This phenomenon is now used universally in magnetic recording to identify various impairments of the magnetic system.

In our case, the blip phenomenon is due to circuit-driver nonlinearity somewhere in the test system. Analysis shows that this imperfection (if so it may be called) is due to a non-linearity in the driver circuit in the performance analyzer. Close examination of the output differential pulses show an asymmetry in the rise times (as compared to the fall times) of the output driver. As a result, a positive-going pulse looks noticeably different than the opposite of a negative-going pulse. This difference is by definition a nonlinear effect.

It has been shown that the PRBS phenomenon is related to the Volterra-series nonlinear-system identification technique for discrete-time systems [4]. This property is another example of the apparently endless set of practical and useful features of a PRBS.

To an excellent approximation, the transmissions lines are natively linear, and (in principle) the choice of excitation waveform is immaterial. However, circuit drivers, receivers, digital-to-analog converters, equalizers, analog-to-digital converters, and other elements of a complex analog communication system, are not natively linear. In such a case, analysis using linear assumptions, under conditions of PRBS excitation yields (for no additional work), nonlinear system analysis capability.

## Comparison to Theory

We now present experimental PRBS results gathered from several different types of transmission lines, and compare the results to expectations from our RLGC model. We found that a constant loss tangent assumption was adequate at lower frequencies (~2.5 Gbits/second excitation) but completely inadequate at 40 Gbits/sec excitation. We concluded that the series losses were modeled adequately by the standard form at all frequencies tested, except for any frequency using a bilayer-conductor transmission system.

### 2.5 Gbits/second Excitation Results

We applied the preceding techniques to a PCB transmission line. We designed and procured a passive printed circuit board (PCB) using GETEK ™ dielectric materials. This board was fabricated to verify the accuracy of the RLGC model described above.

Manual RLGC-model fitting procedures produced model results which compared quite favorably to the experimentally-measured results at 2.5 Gbits/second. These results are compared, for four different lengths of the transmission line, in Figure 6. All model parameters are frozen at the given values (except, of course, for the transmission-line length) for all model results on this graph. We used a 6-inch version of the lines as the de-embedding reference. The important model parameters for the GETEK materials are:

$$L_\infty = 378 \text{ nH/m}, \ C = 117 \text{ pF/m}, \ \tan\delta = 0.011, \sigma = 5.45E7, S/\eta = 5 \text{ mils}. \quad (1.24)$$

We tested several lengths of 100 Ohm Skewclear Infiniband 12X cable from Amphenol/SpectraStrip ™. We merely used the advertised (nominal) odd-mode impedance and velocity numbers to calculate the odd-mode reactive parameters.

The model parameters for the SKEWCLEAR cable are:

$$L_\infty = 377 \text{ nH/m}, \ C = 37.7 \text{ pF/m}, \ \tan\delta = 0.0001, \sigma = 6.0E7, S/\eta = 17 \text{mils}. \quad (1.25)$$

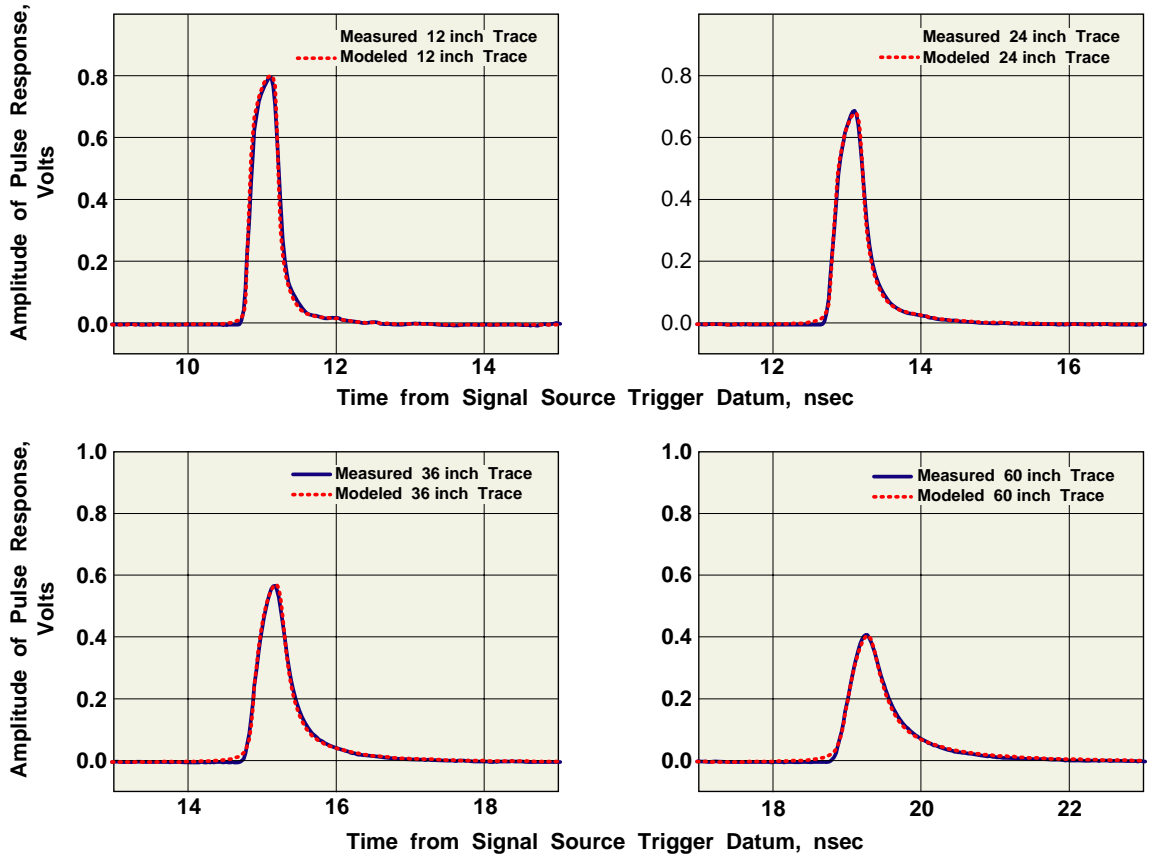

**Figure 6: Experimental Comparison of GETEK PCB to RLGC Model in Odd Mode with Parameters** $L_\infty = 378$ nH/m, $C = 117$ pF/m, $\tan\delta = 0.011, \sigma = 5.45E7, S/\eta = 5$ mils. **(18334)**

Figure 7 shows the modeled and actual pulse responses for several lengths of cable.   In this case, the cable connectors were clearly a significant factor in the test system response as is seen by the ringing in the response tail.   However, the de-embedding method The reference measurement (shown in the first panel of Figure 7, used as input to the numerical model, was taken from a relatively short, 1 meter cable.

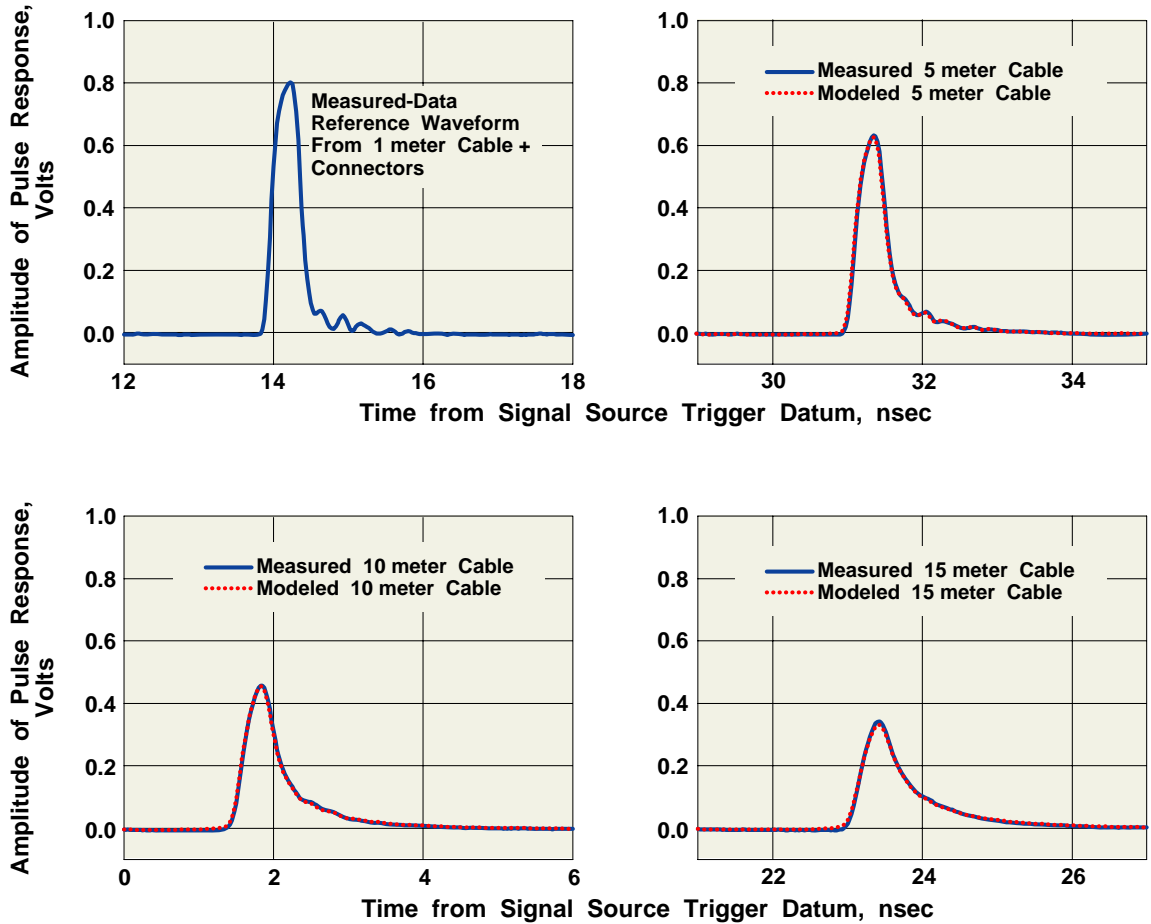

**Figure 7: Experimental Comparison of SkewClear Cable to RLGC Model using Parameters**
$L_\infty = 377$ nH/m, $C = 37.7$ pF/m, $\tan \delta = 0.0001, \sigma = 6.0E7, S/\eta = 17$ mils. (**18324**)

We also analyzed a Gore EyeOpener Plus cable (26 AWG), which was optimized by Gore for operation at 2.5 Gbits/second. The experimental data was not fitted well with the traditional model for surface impedance. We assumed that the cable was built with stratified signal conductors and were able to fit the data much better using equation (1.10) for the surface impedance. The fitted parameters for EyeOpener cable are:

$$L_\infty = 387 \text{ nH/m}, C = 37.7 \text{ pF/m, } \tan \delta = 0.0004, \sigma_1 = 6E7, \sigma_2 = 1E7,$$
$$\tau_1 = 0.115 \text{ mil}, S/\eta = 21 \text{ mils}. \qquad (1.26)$$

The effective conductivity of the core conductor is so low, for a metal, that the material is likely to be magnetic. The effective conductivity of the bulk material is "reduced" by a factor equal to the relative permeability of the material in our model.

Figure 8 shows the modeled and actual pulse responses for two lengths of the Gore EyeOpener Plus cable. The reference measurement, used as numerical model input, was taken from a relatively short, 1 meter cable. Because little or no geometry or composition information was available for the cables, we do not claim that the fit physical parameters truly reflect the physical cable characteristics. We do claim that this combination of parameters adequately describes the differential behavior of the cables over a significant range of cable lengths.

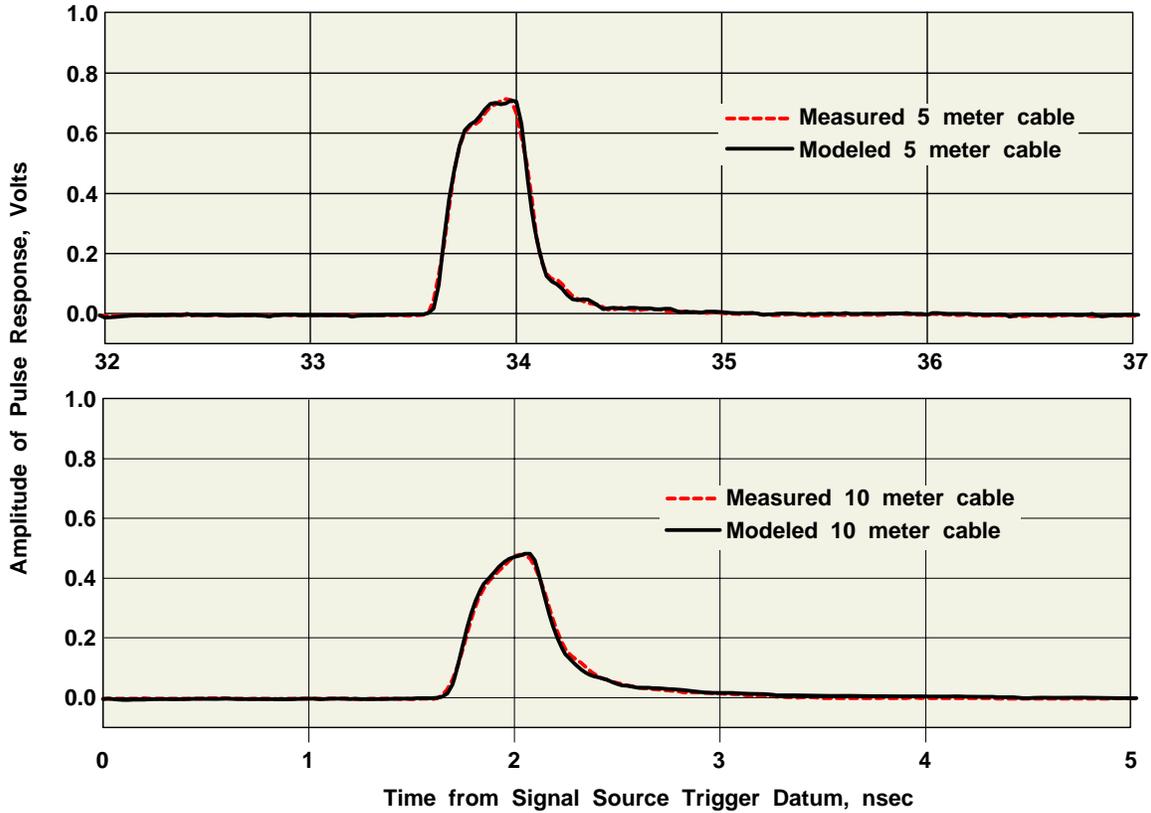

**Figure 8: Experimental Comparison of EyeOpener Cable to RLGC Model using Parameters** $L_\infty = 387$ nH/m, $C = 37.7$ pF/m, $\tan\delta = 0.0004, \sigma_1 = 6E7, \sigma_2 = 1E6, \tau_1 = 0.115$ mil, $S/\eta = 21$ mils. **(18446)**

We conclude that the RLGC model, with appropriate extensions when necessary, accurately describes a variety of cable transmission lines and PCB lines at 2.5 Gbits/second excitation.

## 40 Gbits/second Excitation

We shall now present the time-domain analysis with data taken at 40 Gbits/second excitation. While (in principle) an infinitely-sharp rise time at 2.5 Gbits/second will also excite very high frequencies which will then be identified with by our method even at low excitation frequency, it's also true that equipment made to produce 40 Gbits/second data will have significantly smaller rise times and that of equipment made to generate a PRBS at lower frequencies. Using such equipment is therefore experimentally better if high frequencies are to be accurately modeled by this method.

In Figure 9, we show comparisons for model versus actual pulse responses for various lengths of a PCB trace made using Nelco 1300 ™ material. In this case we found much improved experimental fit using a linear loss tangent model. The graphs show quite a bit of attenuation for this type of transmission line at these frequencies (which is not at all surprising as this material is not intended for use up at these frequencies over any reasonable length of trace). We find the fit of these curves to be fairly good, but clearly there exist differences between expected and actual traces which do not exist in the 2.5 Gbit/second data. The frequency content of these waveforms is only significant up to approximately 10 GHz of bandwidth, so we do not claim to have a model which matches data above this bandwidth.

Model parameters for the 40 Gbits/second Nelco 1300 PCB traces are:

$$L_\infty = 327 \text{ nH/m}, C = 125 \text{ pF/m}, S/\eta = 5.75 \text{ mils}, \tan \delta = 2\text{E-}13\omega, \sigma = 5\text{E}7 \text{ S/m} \quad (1.27)$$

Note that this model uses a loss tangent which is proportional to frequency.

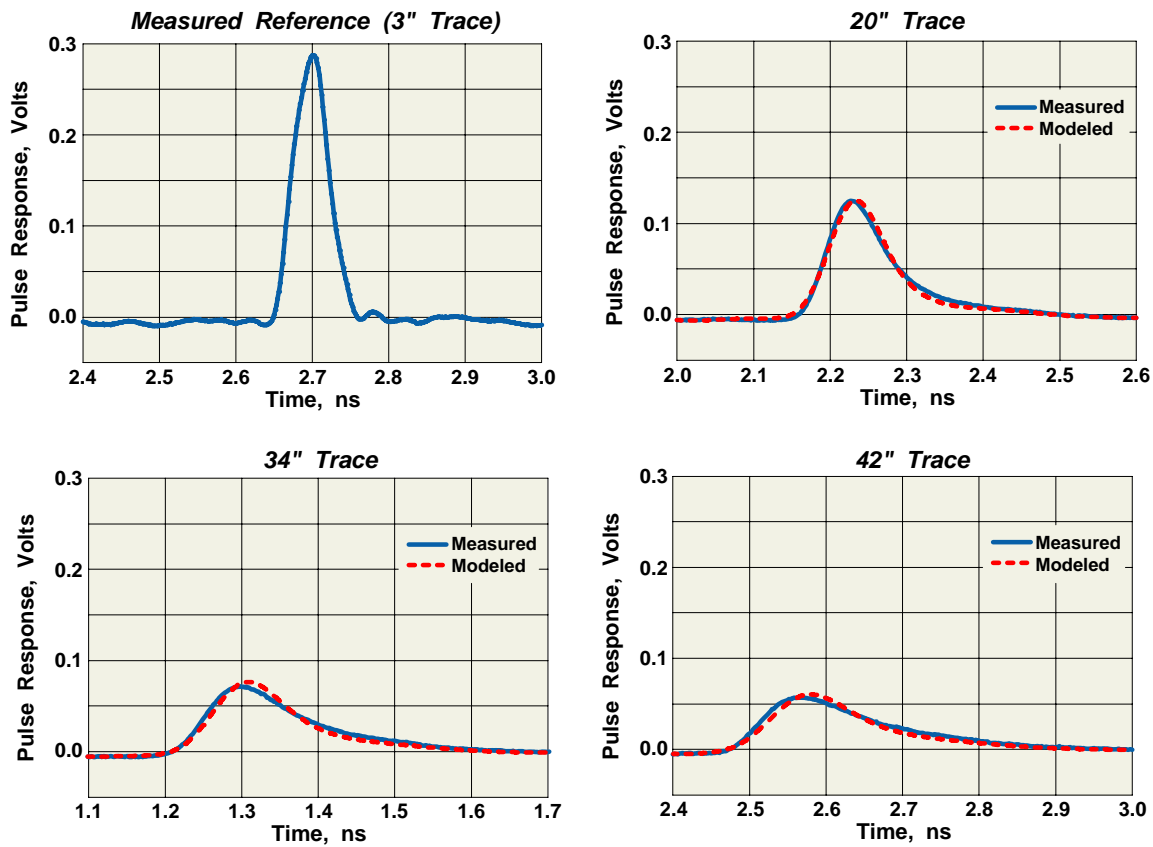

**Figure 9: Experimental Comparison of Nelco 1300 PCB to RLGC Model at 40 Gbits/s using Parameters** $L_\infty = 327$ nH/m, $C = 125$ pF/m, $S/\eta = 5.75$ mils, $\tan \delta = 2\text{E-}13\omega$, $\sigma = 5\text{E}7$ S/m **(20568)**

Finally, in Figure 10 we show the results from data taken at 40 Gbits/second from an advanced version of the SkewClear cable. This version is designed for higher frequencies and includes a modified ground conductor.

The model parameters for the upgraded SkewClear cable are:

$$L_\infty = 451 \text{ nH/m}, C = 40 \text{ pF/m}, S/\eta = 20 \text{ mils}, \tan \delta = 1\text{E-}13\omega, \sigma = 5\text{E}7 \text{ S/m} \quad (1.28)$$

Note that again the loss tangent model is linear in frequency.

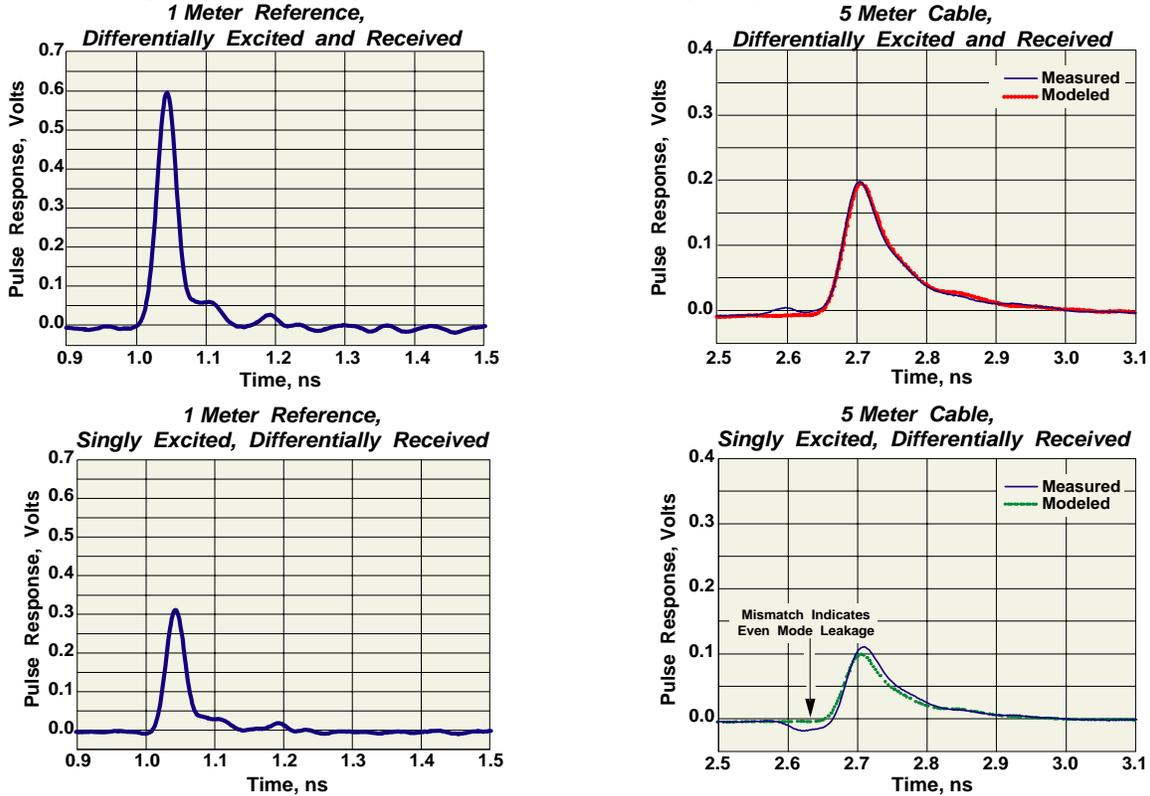

**Figure 10: Experimental Comparison of Upgraded SkewClear Cable to RLGC model at 40Gb/s under Single Ended and Differential Excitation Conditions using Model Parameters**
$L_\infty = 451 \text{ nH/m}, C = 40 \text{ pF/m}, S/\eta = 20 \text{ mils}, \tan \delta = 1\text{E-}13\omega, \sigma = 5\text{E}7 \text{ S/m}$ **(20571)**

In this case, the fit is nearly as excellent as for the 2.5 GHz case, even though the significant frequencies in this waveform extend up to approximately 20 GHz. The usual model comparison (in the upper right graph) shows excellent fit except for the area around 100 pS before the main pulse, which shows a relatively small blip of apparently unknown origin.

We shall take this opportunity to show the types of conclusions one may draw from investigations using time domain techniques. We have shown similar data on the lower portion of the figure except in this case the excitation is provided only on one side of the differential cable, with the other input terminated. Theoretically, or ideally, this condition should produce exactly half the differential signal at the output terminals (and indeed this is very well approximated for the 1 meter reference signals). However, the singly-excited model comparison is rather weak, though (evidently) the errors apparently nearly cancel when the system is both differentially excited and differentially received (the upper right case). The simplest explanation for this behavior is a mode leakage between even and odd modes down the length of the line; and indeed, we have shown by similar methods that the even mode signal has a slightly higher velocity than does the odd mode, such that the even mode shows up about 100 pS before the odd mode at a distance of 5 meters.

## Conclusions

We have developed a time-domain method for experimental verification of an RLGC model, and showed adequate experimental fit over a wide variety of transmission line types up to approximately 10 GHz in bandwidth. We found that, depending on the frequency and transmission line type, specific extensions were required to adequately explain laboratory behavior. We found that modifications to both the dielectric (shunt) loss model and the resistive (series) loss models were necessary to adequately cover all cases.

We have shown that, despite a fairly substantial list of assumptions, a PRBS time-domain method can be used in a de-embedding fashion to remove many experimental impediments. Indeed, the PRBS technique can be used to easily identify and quantify nonlinear effects.

We proposed an extremely efficient method for implementing the PRBS identification process, which can either speed up offline computation when the waveform is large, or, can be used in hardware to directly implement the method in real-time so long as an appropriately complex ADC is available.

We note that the proposed technique is similar to (in fact, it is a simplification of) a generalized TDT method, possessing a similar set of advantages and disadvantages when compared to frequency-domain methods. For the price of software processing and a real time scope, this method can be used to turn a bit-error-rate test system into a reasonable system identification station.

While we have shown laboratory data indicating the utility of this measurement method using high-quality lab equipment, we believe that this type of analysis will become most attractive in self-test applications in product hardware. As the analog-to-digital converters (ADCs) in a SERDES migrate from traditional 1-bit comparators to those required to support advanced signaling and detection schemes, we expect that methods such as those described in this paper will be used to advance the operational and diagnostic capabilities of the overall system, as they have in other systems [12].

## Acknowledgements

The authors would like to thank Eric Hanlon and Devon Post for providing test boards and cables for this activity; Jason Prairie for his data-taking expertise; Patrick Zabinski for many instructive and seminal conversations; and Steve Richardson, Elaine Doherty, and Deanna Jensen for generating the graphics for this report.## References